\documentclass[runningheads,a4paper]{llncs}

\usepackage{amssymb}
\usepackage{graphicx}
\usepackage{color}
\usepackage{ulem}

\usepackage{url}
\urldef{\mailsa}\path|{a.seyfried, a.portz}@fz-juelich.de|
\urldef{\mailsb}\path|as@thp.uni-koeln.de|
\newcommand{\keywords}[1]{\par\addvspace\baselineskip
\noindent\keywordname\enspace\ignorespaces#1}

\begin{document}

\mainmatter  % start of an individual contribution

\title{Phase coexistence in congested states of pedestrian dynamics}

\titlerunning{Phase coexistence in congested states of pedestrian
dynamics} 

\author{Armin Seyfried\inst{1}, Andrea Portz\inst{1}%
%\thanks{Please note that the LNCS Editorial assumes that all authors have used
%the western naming convention, with given names preceding surnames. This determines
%the structure of the names in the running heads and the author index.}%
\and Andreas Schadschneider\inst{2}}
\authorrunning{A. Seyfried, A. Portz and A. Schadschneider}
\institute{J\"ulich Supercomputing Centre, Forschungszentrum J\"ulich, 52425
  J\"ulich, Germany\\ 
\mailsa\\
\and
Institut f\"ur Theoretische  Physik, Universit\"at zu K\"oln, 50937 K\"oln,
Germany\\ 
\mailsb\\
}

\toctitle{Lecture Notes in Computer Science}
\tocauthor{Authors' Instructions}
\maketitle

\begin{abstract}
  Experimental results for congested pedestrian traffic are presented.
  For data analysis we apply a method providing measurements on an
  individual scale. The resulting velocity-density relation shows a
  coexistence of moving and stopping states revealing the
  complex structure of pedestrian fundamental diagrams and supporting new
  insights into the characteristics of pedestrian congestions.
  Furthermore we introduce a model similar to event driven approaches. The
  velocity-density relation as well as the phase separation is
  reproduced. Variation of the parameter distribution indicates that the
  diversity of pedestrians is crucial for phase separation.  
  
\end{abstract}

%%%%%%%%%%%%%%%%%%%%%%%%%%%%%%%%%%%%%%%%%%%%%%%%%%%%%%%%%%%%%%%%%%%%%%%%

\section{Introduction}

The velocity-density relation is one of the most important
characteristics for the transport properties of any traffic system.
For pedestrian traffic there is currently no consensus even about the
principle shape of this relation which is reflected e.g.\ in
conflicting recommendations in various handbooks and guidelines
\cite{Schadschneider2009}.  Discrepancies occur in particular in the
high-density regime which is also the most relevant for applications
in safety analysis like evacuations or mass events.  At high densities
stop-and-go waves occur indicating overcrowding and potentially
initiating dangerous situations due to stumbling etc.  However, the
densities where the flow breaks down due to congestion ranges from
densities of $\rho_{\rm max}=3.8$~m$^{-2}$ to $\rho_{\rm max}=10$~m$^{-2}$ 
\cite{Schadschneider2009}.  This large variation in values for 
$\rho_{\rm max}$ reported in the literature is partly due to insufficient 
methods of data capturing and data analysis. In previous experimental 
studies, different kinds of measurement methods are used and often a 
mixture of time and space averages are realized. 
Especially in the case of spatial and temporal inhomogeneities the
choice of the measurement method and the type of averaging have a
substantial influence on the results \cite{Seyfried2010}.

Up to now, congested states in pedestrian dynamics have not been
analyzed in much detail. This is in contrast to vehicular traffic
where the congested phase is well-investigated, both empirically
and theoretically \cite{ChowdhurySS,SCNBook}. 

In this contribution we show that even improved classical measurement
methods using high precision trajectories but basing on mean values of
density and velocity fail to resolve important characteristics of
congested states. For a thorough analysis of pedestrian congestion we
apply a new method enabling measurements on the scale of single
pedestrians.

%%%%%%%%%%%%%%%%%%%%%%%%%%%%%%%%%%%%%%%%%%%%%%%%%%%%%%%%%%%%%%%%%%%%%%%%

\section{Experimental data}

For our investigation we use data from experiments performed in 2006
in the wardroom of Bergische Kaserne D\"usseldorf with a test group of
up to $N=70$ soldiers. The length of the circular system was about 26~m, 
with a $l = 4$~m long measurement section.  Detailed information about the
experimental setup and data capturing providing trajectories of high accuracy
($|x_{\rm err}| \leq 0.02$~m) is given in \cite{Seyfried2010,Boltes2010}. 

\begin{figure}
  \centering
  \includegraphics[width=0.45\textwidth]{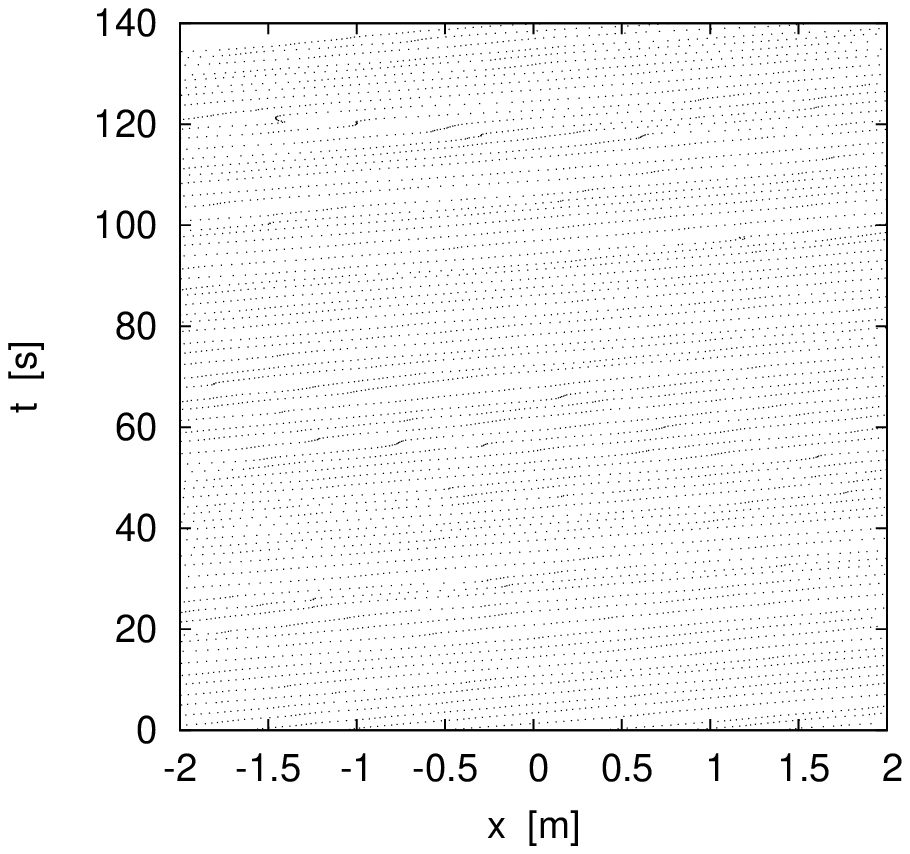}
  \quad
  \includegraphics[width=0.45\textwidth]{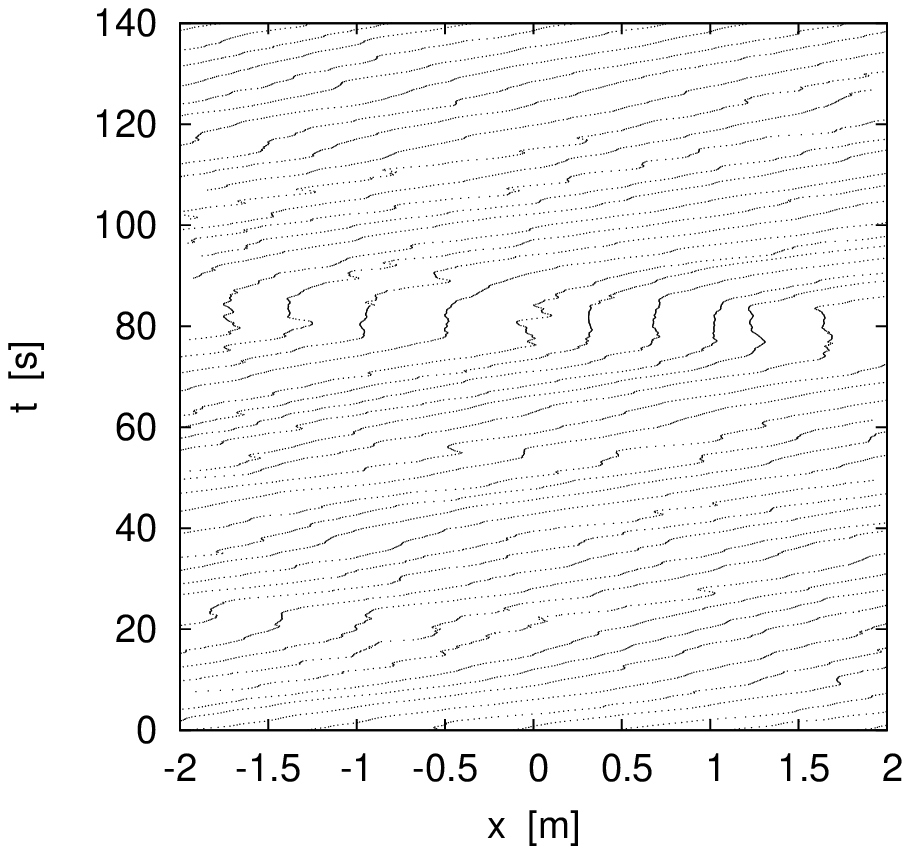}
  \includegraphics[width=0.45\textwidth]{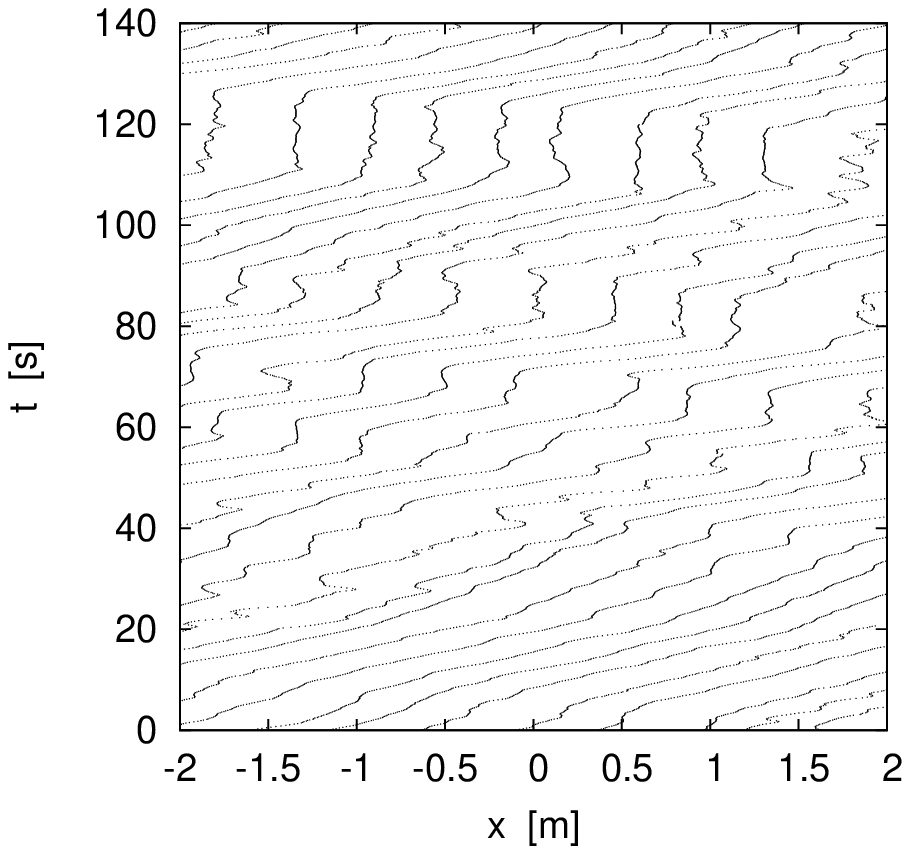}
  \quad
  \includegraphics[width=0.45\textwidth]{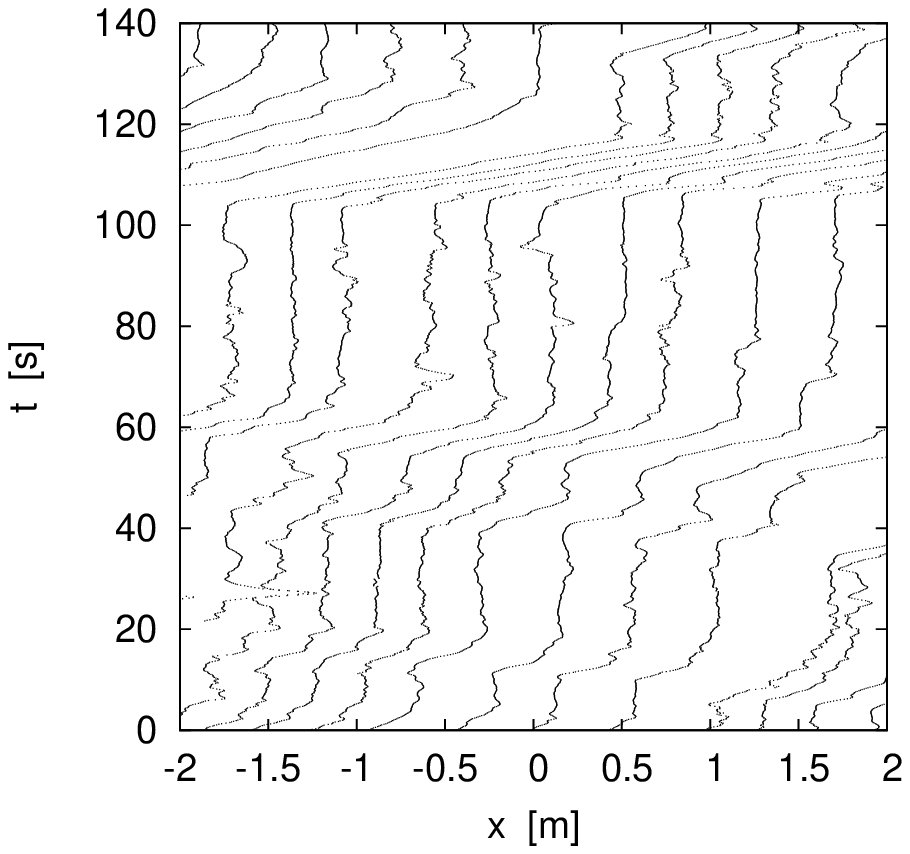}
  \caption{Trajectories for the runs with $N=39,56,62$ and $70$ (left to right,
    top to bottom). With increasing density the occurrence of stop-and-go waves 
    accumulate.}
  \label{fig:traj}
\end{figure}

In Fig.~\ref{fig:traj} the $x$-component of trajectories is plotted
against time. For the extraction of the trajectories, the pedestrians'
heads were marked and tracked. Backward movement leading to negative
velocities is caused by head movement of the pedestrians during a
standstill. Inhomogeneities in the trajectories increase with
increasing density. As in vehicular traffic, jam waves
propagating opposite to the movement direction (upstream) occur at
higher densities.

Stopping is first observed during the runs with $N=45$ pedestrians, at
$70$ pedestrians they can hardly move forward.  Macroscopically one
observes separation into a stopping area and an area where
pedestrians walk slowly.  In the following we analyze how macroscopic
measurements blur this phase separation and apply a
technique introduced in \cite{Steffen2010} enabling a measurement of the
fundamental diagram on a `microscopic' scale.

\subsection{Macroscopic measurement}

Speed $v_i$ of pedestrian $i$ and the associated density $\rho_i$ are
calculated using the entrance and exit times $t_i^{\rm in}$ and
$t_i^{\rm out}$ into and out of a measurement section of 
length $l_m= 2$~m,

\begin{equation}
  v_i = \frac{l_{m}}{t_{i}^{\rm out}-t_{i}^{\rm in}}, \qquad 
  \rho_i = \frac{1}{t_{i}^{\rm out}-t_{i}^{\rm in}} \int_{t_i^{\rm
      in}}^{t_{i}^{\rm out}} \rho(t) \, dt
  \quad \mbox{with} \quad 
  \rho(t) = \frac{\sum{\Theta_i(t)}}{l_m } \,.
\end{equation}

\begin{figure}
  \centering
  \includegraphics[width=0.46\textwidth]{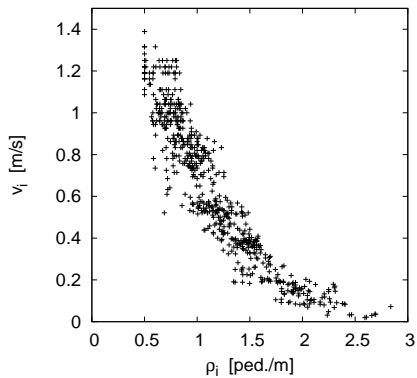}
  \caption{Velocity-density relation using an improved macroscopic
    measurement method. There is no indication of phase separation since
    stopping states  ($v\approx 0$) occur only at large densities where
    no moving states ($v>0$) are observable.}
  \label{fig:fdmacro}
\end{figure}

The speed $v_i$ is a mean value over the space-time interval $\Delta
t_i=t_i^{\rm out}-t_i^{\rm in}$ and  $\Delta x=l_m$. 
By integration over the instantaneous density $\rho(t)$ the density is
assigned to the same space-time interval. To reduce the fluctuations
of $\rho(t)$ we use the quantity $\Theta_i(t)$, introduced in
\cite{Seyfried2005}, which measures the fraction of space between
pedestrians $i$ and $i+1$ that is inside the measurement area.

Results of the macroscopic measurement method are shown in
Fig.~\ref{fig:fdmacro}. In comparison to method B introduced in
\cite{Seyfried2010} (see Fig.~6 of \cite{Seyfried2010} which
  uses data based on the same trajectories as this study) the scatter of the
data is reduced due to an improved density definition and a better assignment
of density and velocity. However the resulting velocity-density relation does
not allow to identify phase-separated states although these are clearly
visible in the trajectories. 

\subsection{Microscopic measurement}

To identify phase separated states we determine the velocity-density
relation on the scale of single pedestrians. This can be achieved by
the Voronoi density method \cite{Steffen2010}. In one dimension a
Voronoi cell is bounded by the midpoints $z_i= ({x_{i+1}+x_i})/{2}$
of the pedestrian positions $x_i$ and $x_{i+1}$.  With the length
$L_i=z_{i}-z_{i-1}$ of the Voronoi cell corresponding to pedestrian
$i$ and $\Delta t=0.5$ s we define the instantaneous velocity and
density by

\begin{equation}
  v'_i(t) =  \frac{x_i(t+\Delta t/2)-x_i(t-\Delta t/2)}{\Delta t}, 
  \; \rho'_i(x) =  \left\{\begin{array}{r@{\quad:\quad}l} 
  {1}/{L_{i}} & x \in [z_{i},z_{i+1}[\\ 
      0 & \mbox{otherwise} 
\end{array} \right. \,.
\end{equation}

\begin{figure}
\centering
\includegraphics[width=0.48\textwidth]{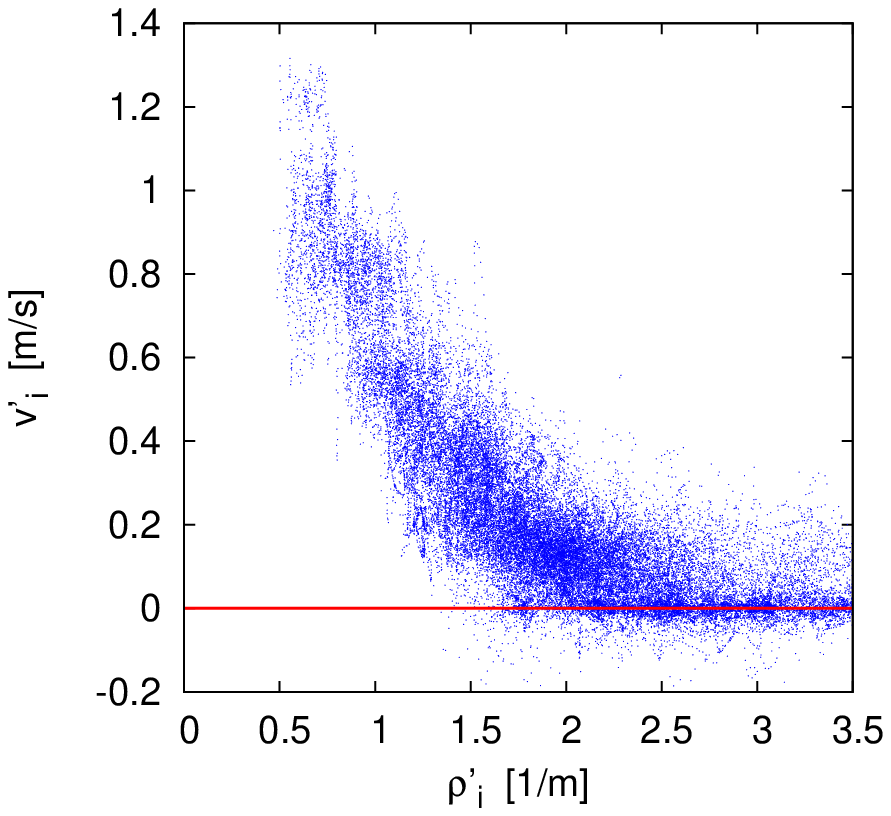}
\quad
\includegraphics[width=0.48\textwidth]{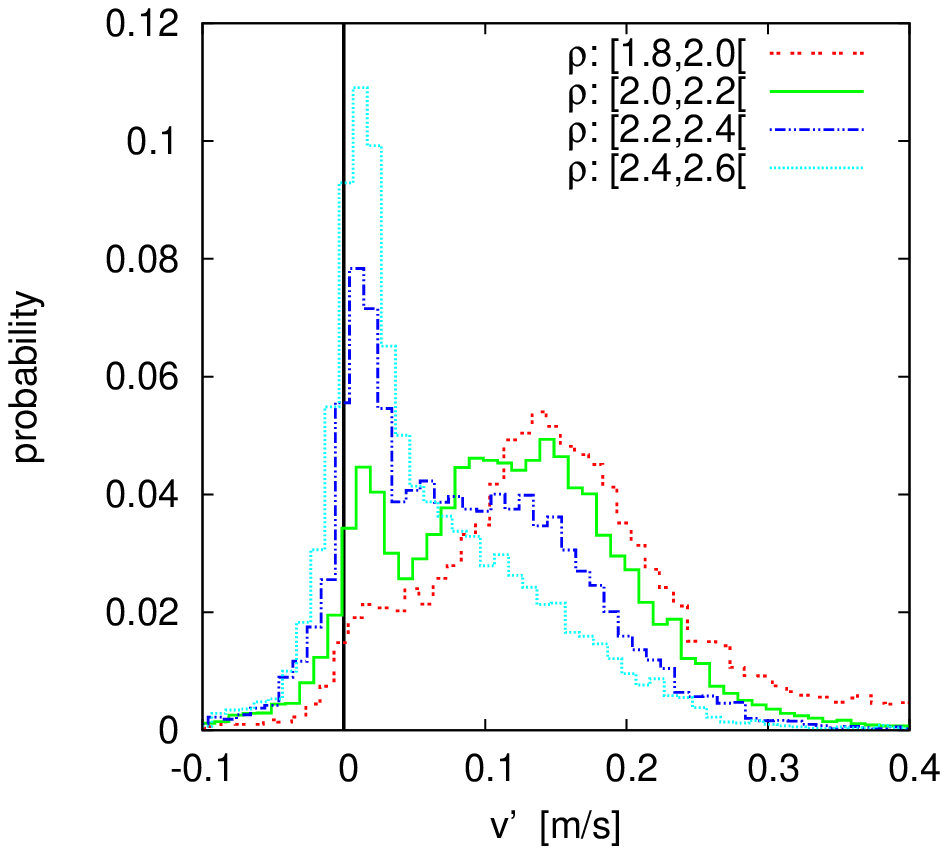}
\caption{{\bf Left:} Velocity-density relation on an individual scale. 
  At each density $\rho' > 1.5~$m$^{-1}$ both stopping and 
  moving states are observable. {\bf Right:} Probability distribution of the
  velocities for different density intervals. The double peak structure
  indicates the coexistence of moving and stopping states.
  The height of the stopping peak increases with increasing
  density.}  
\label{fig:fdmicro}
\end{figure}

The fundamental diagram based on the Voronoi method is shown on the
left side of Fig.~\ref{fig:fdmicro}. Regular stops occur at densities
higher than 1.5~m$^{-1}$. On the right side of Fig.~\ref{fig:fdmicro}
the distribution of the velocities for fixed densities from
1.8~m$^{-1}$ to 2.6~m$^{-1}$ are shown. There is a continuous change
from a single peak near $v = 0.15~$m/s, to two peaks, to a single peak
near $v= 0~$m/s. The right peak represents the moving phase,
whereas the left peak represents the stopping phase.  At
densities around 2.2~m$^{-1}$ these peaks coexist, indicating phase separation 
into a flowing and a jammed phase.

\subsection{Phase separation in vehicular traffic}

In highway traffic, phase separation into moving and stopping phases typically
occurs when the outflow from a jam is reduced compared to maximal possible
flow in the system. Related phenomena are hysteresis and a non-unique
fundamental diagram. At intermediate densities two different flow values can
be realized. The larger flow, corresponding to a homogeneous state, is
metastable and breaks down due to fluctuations or perturbations (capacity
drop).  The origin of the reduced jam outflow is usually ascribed to the
so-called slow-to-start behaviour (see \cite{ChowdhurySS,SCNBook} and
references therein), i.e.\ an delayed acceleration of stopped vehicles due to
the loss of attention of the drivers etc.   

The structure of the phase-separated states in vehicular traffic is different
from the ones observed here.  For vehicle traffic the stopping phase
corresponds to a jam of maximal density whereas in the moving phase the flow
corresponds to the maximal {\it stable} flow, i.e.\ all vehicles in the moving
phase move at their desired speed. This scenario is density-independent as
increasing  the global density will only increase the length of the stopping
region without reducing the average velocity in the free flow regime. The
probability distribution of the velocities (in a  periodic system) shows a
similar behaviour to that observed in Fig.~\ref{fig:fdmicro}. The position of
the free  flow peak in the case of vehicular traffic is independent of the
density. 

The behaviour observed here for pedestrian dynamics differs slightly from that
described above. The main difference concerns the properties of the moving
regime. Here the  observed average velocities are much smaller than the free
walking speeds. Therefore the two regimes observed in the phase separated 
state are better characterized as ``stopping" and ``slow moving" regimes.

Further empirical studies are necessary to clarify the origin of these
differences. One possible reason are the  different acceleration properties of
vehicles and pedestrians as well as anticipation effects. It also remains to
be seen whether in pedestrian systems phenomena like hysteresis can be
observed. 

%%%%%%%%%%%%%%%%%%%%%%%%%%%%%%%%%%%%%%%%%%%%%%%%%%%%%%%%%%%%%%%%%%%%%%%%

\section{Modeling}

\subsection{Adaptive Velocity Model}

\begin{figure}
\centering
\includegraphics[height=4.5cm]{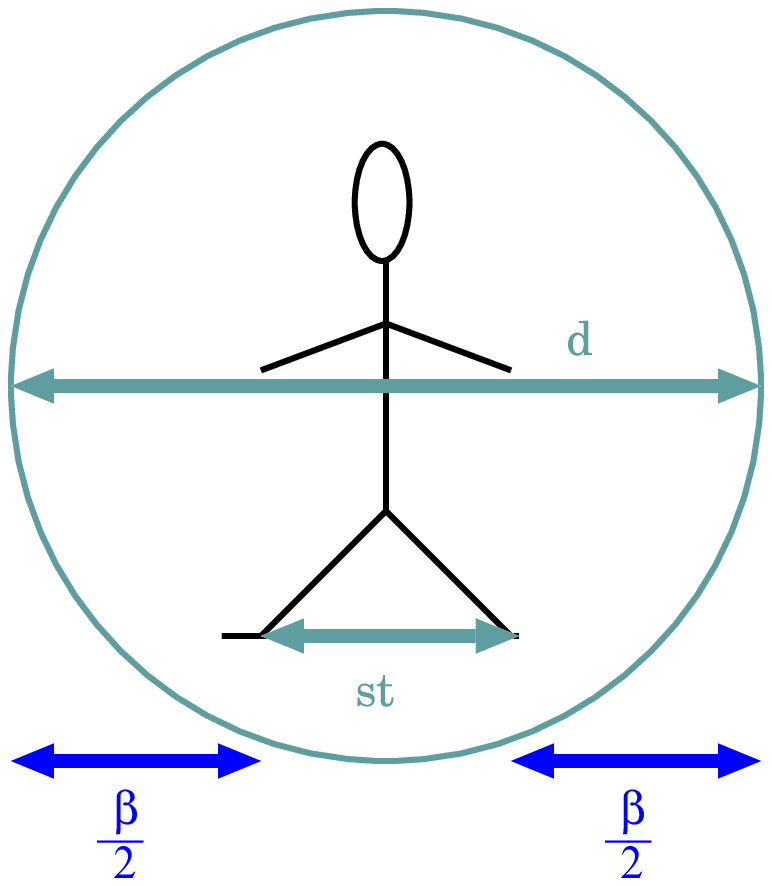}
\includegraphics[height=4.5cm]{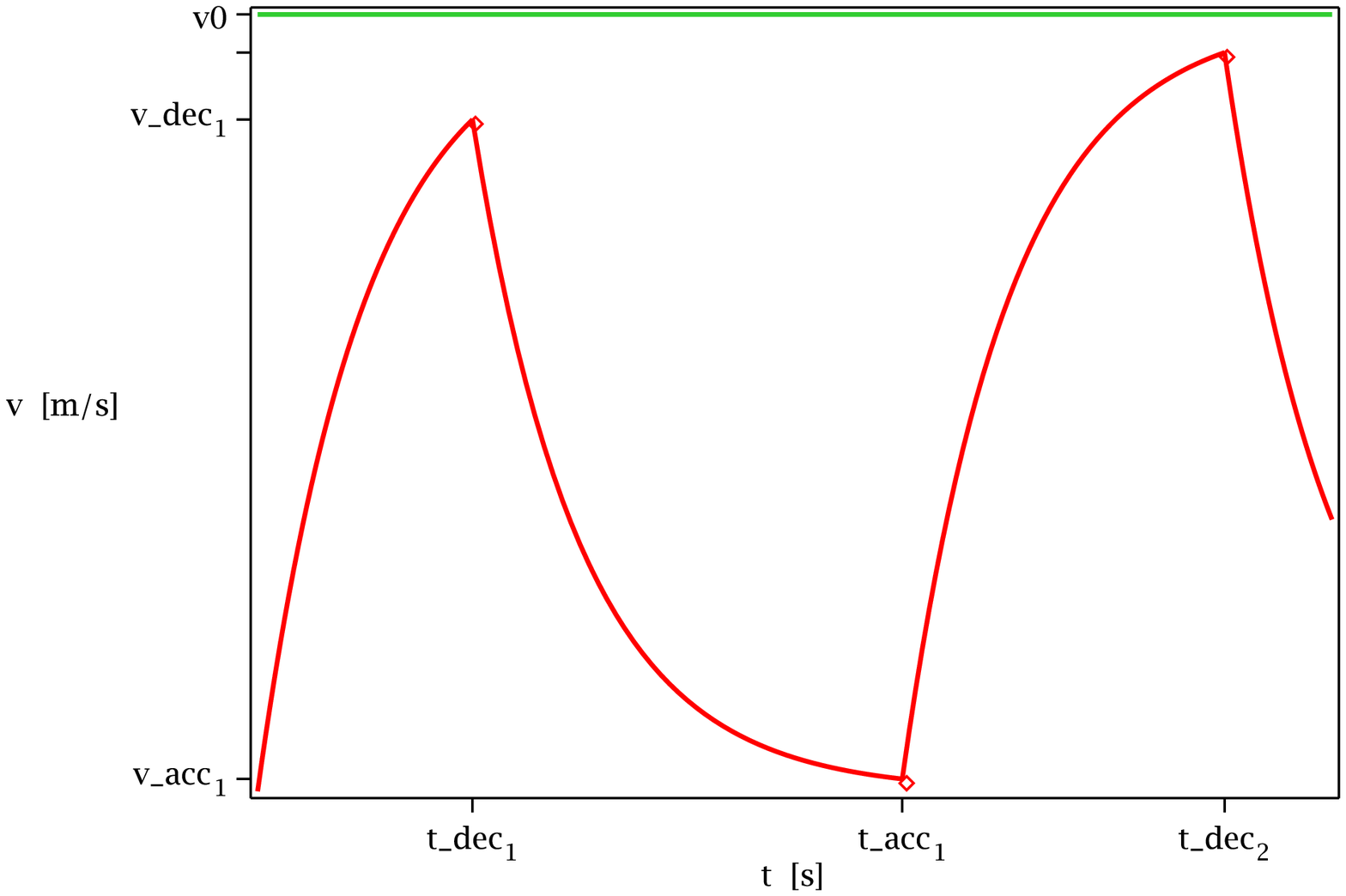}
\caption{\textbf{Left:} Connection between the 
    required space $d$, the step length $st$ and the safety distance
    $\beta$.  \textbf{Right:} The adaptive velocity: acceleration
    until $t_{\rm dec1}$ than deceleration until $t_{\rm acc1}$,
    again acceleration until $t_{\rm dec2}$ and so on.}
\label{fig:paraModell}
\end{figure}

In this section we introduce the adaptive velocity model, which is based on an
event driven approach \cite{Chraibi2010}.  A pedestrian can be in different
states which determine the velocity. A change between these states is called
{\it event}.  The model was derived from force-based models, where the
dynamics of pedestrians are given by the following system of coupled
differential equations 
\begin{equation}
  m_i \frac{dv_i}{dt} = F_i \quad\quad \mbox{with} \quad 
  F_i = F_i^{\rm drv} + F_i^{\rm rep} \quad 
  \mbox{and} \quad \frac{dx_i}{dt}=v_i,
\end{equation}
where $F_i$ is the force acting on pedestrian $i$. The mass is denoted
by $m_i$, the velocity by $v_i$ and the current position by $x_i$.
$F_i$ is split into a repulsive force $F_i^{\rm rep}$ and a driving force
$F_i^{\rm drv}$. The dynamics is regulated by the interrelation between
driving and repulsive forces. In our approach the role of repulsive
forces are replaced by events. The driving force is defined as
\begin{equation}
  F_i^{\rm drv} = \frac{v_i^0-v_i}{\tau_i},
\end{equation}
where $v_i^0$ is the desired speed of a pedestrian and $\tau$ the 
relaxation time of the velocity. By solving the differential equation  
\begin{equation}
  \frac{dv_i}{dt} = F_i^{\rm drv} \quad \Rightarrow \quad v_i(t) 
  = v_i^0+c\exp\left(-\frac{t}{\tau_i}\right), \quad
\mbox{with } c \in \mathbb{R},
\end{equation}
the velocity function is obtained. This is shown in Fig.~\ref{fig:paraModell} 
together with the parameters governing the pedestrians' movement.
In this model pedestrians are treated as bodies with diameter $d$
\cite{Chraibi2010}. The diameter depends linearly on the current velocity and
is equal to the step length $st$ in addition to the safety distance $\beta$ 
\begin{equation}
 d_i(t) = st_i(t) + \beta_i(t)
\end{equation}

Step length and safety distance are introduced to define the rules for the
dynamics of the system. We determine  the model parameters from empirical data
which allows to judge the adequacy of the rules. Based on \cite{Weidmann1993}
the step length  is a linear function of the current velocity with following
parameters:    

\begin{equation}
 st_i(t) = 0.235 m + 0.302 [s] \; v_i(t).
\end{equation}

The required quantities for the safety distance can be specified through
empirical data of the fundamental diagram $\bar{d}_i = 1/\rho = 0.36+1.06\;v$,
see  \cite{Seyfried2005}. With these experimental results the previous
equations can be summarized to 
\begin{equation}
  \beta_i(t) = d_i(t)-st_i(t) = a_i + b_i \;  v_i(t) 
\end{equation}
with $a_i=0.125$ m and $b_i=0.758$ s. No free model parameter remain with
these specifications. In the following we describe the rules for the 
movement. A pedestrian accelerates to the desired  velocity $v_i^0$ until the
distance $\Delta x_{i,i+1}$ to the pedestrian in front is smaller than the
safety distance. From this time on, he/she decelerates until the distance is
larger than the safety distance. To guarantee a minimal
  volume exclusion, case ``collision'' is included, in which 
the pedestrians are too close to each other and have to stop.
Via $\Delta x_{i,i+1}$, $d_i$ and $\beta_i$ the velocity function for the
states deceleration (dec.),  acceleration (acc.) and collision (coll.) can be
defined, see Eq. \ref{eq:vel}: 
\begin{equation}
 v_i(t)=\left\{
\begin{array}{ccll}
  v_{\rm dec}\, \exp\left(-\frac{t-t_{\rm dec}}{\tau}\right)
    & \quad \mbox{for} \quad & \Delta x_{i,i+1}-\delta_i(t) \le 0 & \quad
  \mbox{(dec.)} \\
\\
  v_i^0(1-\exp\left(-\frac{t-t_0}{\tau}\right) 
    & \quad \mbox{for} \quad &  \Delta x_{i,i+1}-\delta_i(t) > 0 & \quad
  \mbox{(acc.)} \\\
\\
  0 & \quad \mbox{for} \quad &  \Delta x_{i,i+1}-\delta_i(t) \le -\beta_i(t)/2
  & \quad \mbox{(coll.)} \label{eq:vel}\\
\end{array}\right.
\end{equation}
where $\delta_i(t) = (d_i(t)+d_{i+1}(t))/2$ is the distance between the
centers of both pedestrians. The current velocity $v_i(t)$ of an pedestrian $i$ depends
on his/her state. $t_{\rm dec}$ denominates the point in time where a change from
acceleration to deceleration takes place. Conversely $t_{\rm acc}$ is the change from
deceleration to acceleration. $v_{\rm dec}=v(t_{\rm dec})$ and $v_{\rm
  acc}=v(t_{\rm acc})$ are defined accordingly. At the beginning $t_0 = 0$ s
with a change from acceleration to deceleration a new calculation of $t_0$ is
necessary:  

\begin{equation}
  t_0 = t_{acc} + \ln \left( 1-\frac{v_{\rm acc}}{v_0} \exp\left(\frac{-t_{\rm
      acc}-t_{\rm dec}}{\tau}\right)\right)
\end{equation}

The discreteness of the time step could lead to configurations where
overlapping occurs. To ensure good computational performance for high
densities, no events are explicitly calculated. Instead in each time step, 
it is checked whether an event has taken place and  $t_{\rm dec}$, $t_{\rm acc}$
or $t_{\rm coll}$ are set to $t$ accordingly. To avoid too large
interpenetration of pedestrians and to implement a reaction time in a
realistic size we choose $\Delta t = 0.05$ s. To guarantee a parallel update
a recursive procedure is necessary:  Each  
person is advanced one time step according to Eq.~\ref{eq:vel}. If after this
step a pedestrian is in a different state because of the new distance to the
pedestrian in front, the velocity is set according to this state.  Then the
state of the next following person is reexamined. If the state is still valid
the update is completed. Otherwise, the velocity is calculated again.

\subsection{Model validation and influence of individual differences}

\begin{figure}
  \centering
  \includegraphics[width=0.32\textwidth]{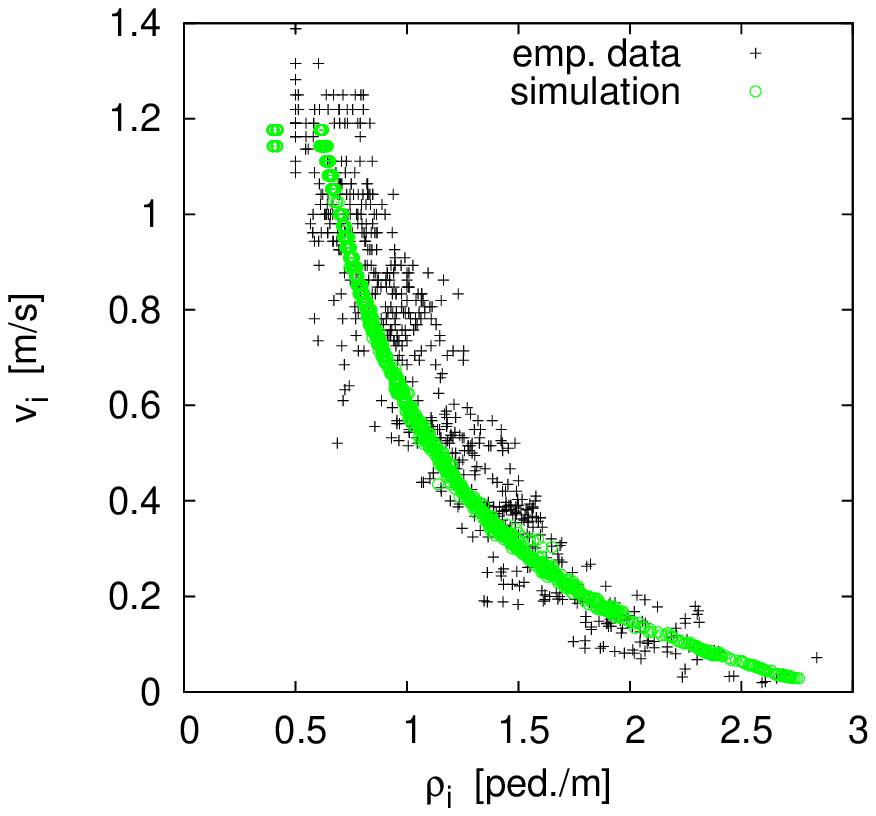}
  \includegraphics[width=0.32\textwidth]{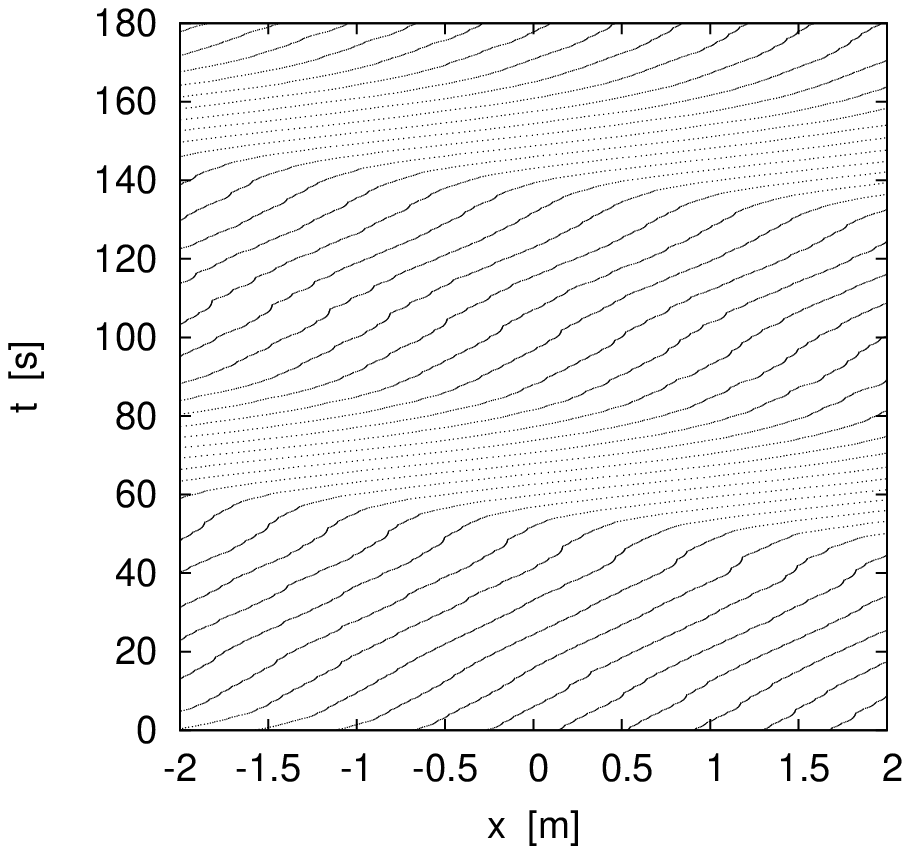}
  \includegraphics[width=0.32\textwidth]{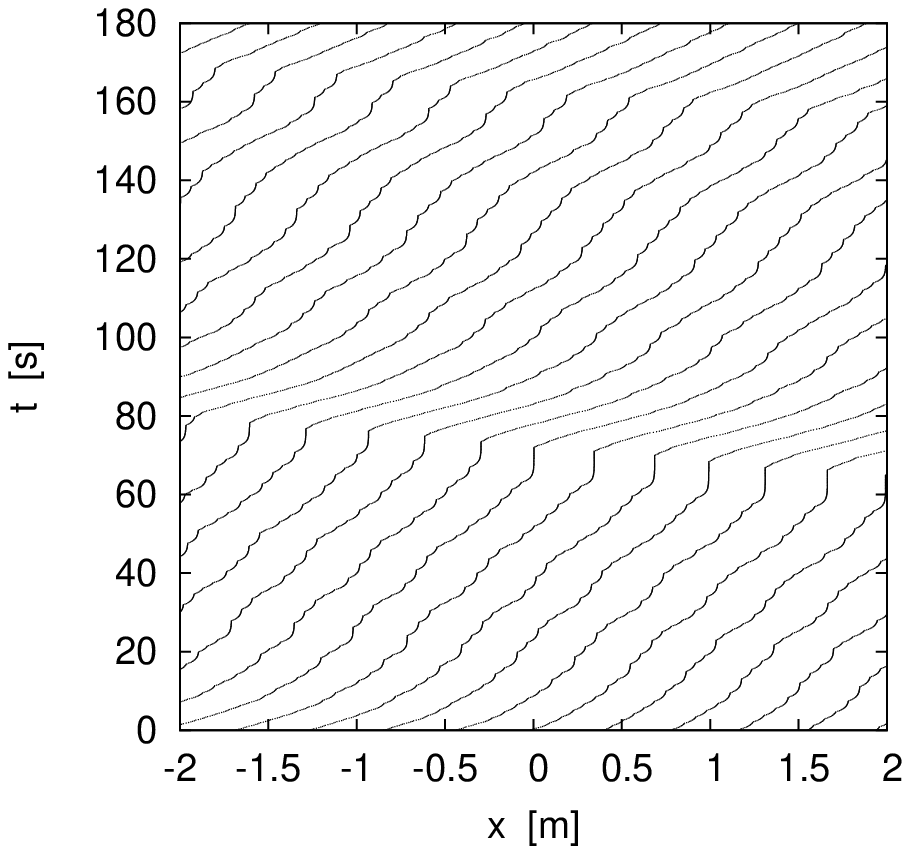}
  \caption{Validation of the modeled fundamental diagram and trajectories with
    same parameter $a_i$, $b_i$ and $\tau_i$ for all pedestrians {\bf Left:}
    Comparison of fundamental diagrams of modeled and empirical data. {\bf
      Middle:} Trajectories for $N=62$ (model). {\bf Right:}  Trajectories for
    $N=70$ (model).}
  \label{fig:simohne}
\end{figure}

\begin{figure}
  \centering
  \includegraphics[width=0.32\textwidth]{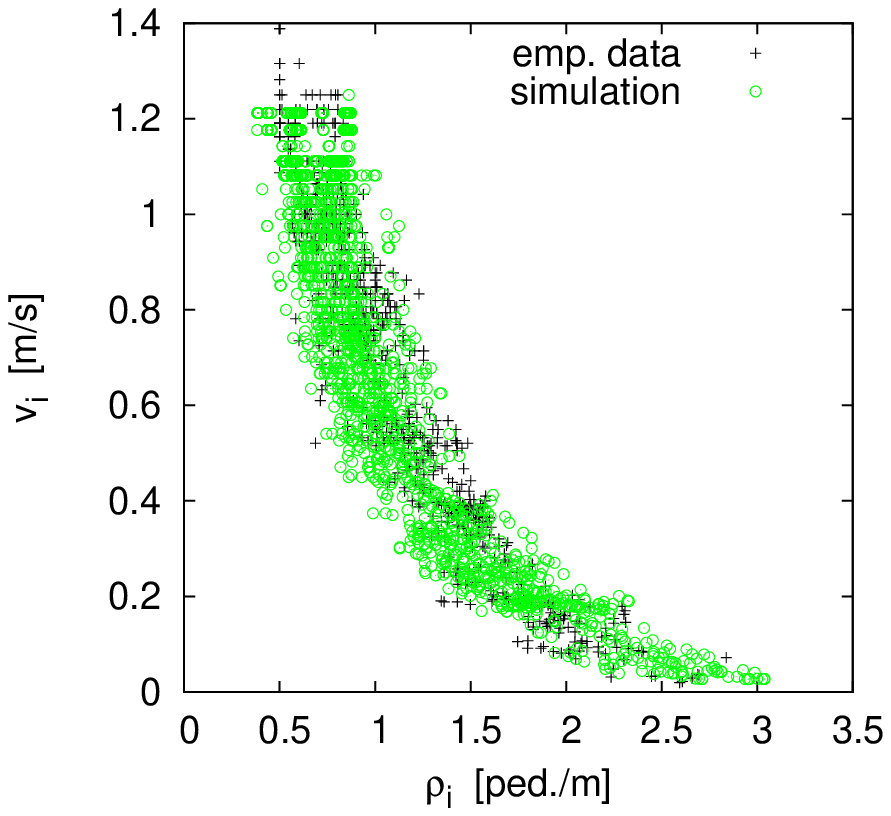}
  \includegraphics[width=0.32\textwidth]{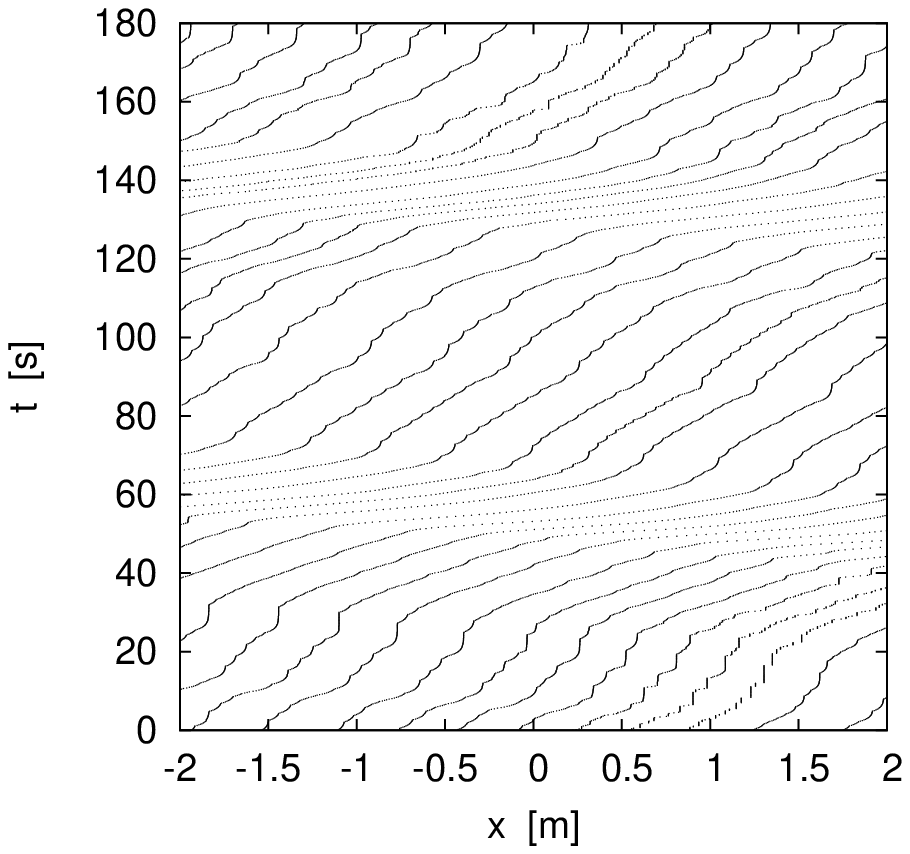}
  \includegraphics[width=0.32\textwidth]{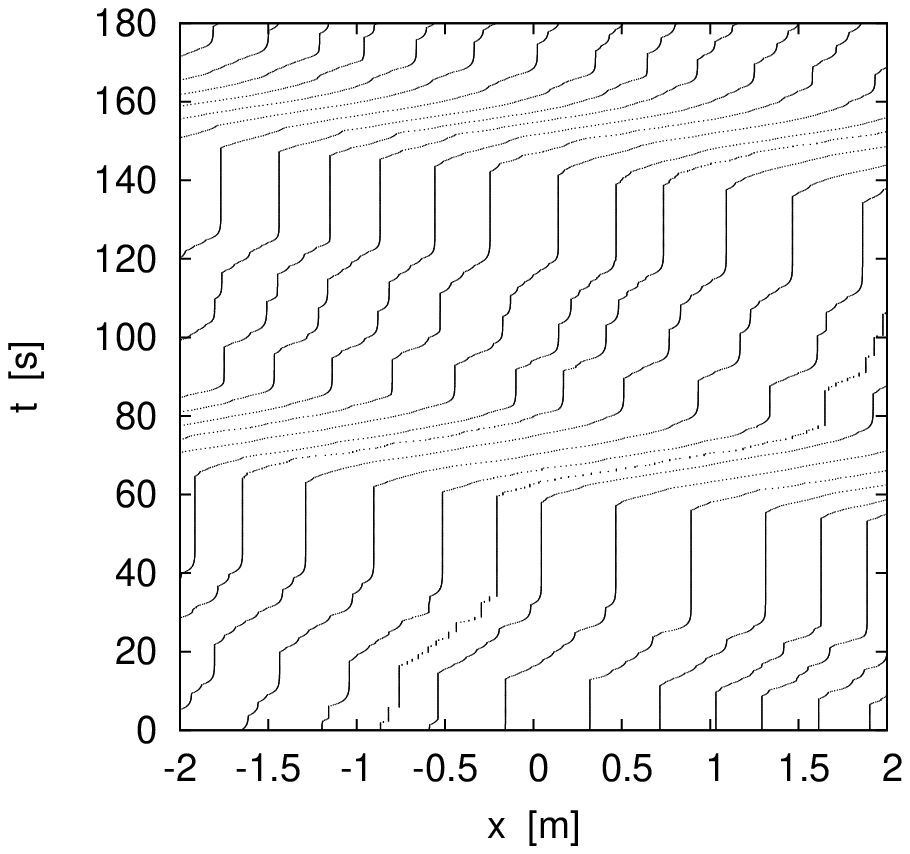}
  \caption{Validation of the modeled fundamental diagram 
    and trajectories, with normal distributed individual parameter. {\bf Left:}
    Comparison of fundamental diagrams of modeled and empirical data. {\bf
      Middle:} Trajectories for $N=62$ (model). {\bf Right:} Trajectories for
    $N=70$ (model).}
  \label{fig:simmit}
\end{figure}

In the following we face model results with experimental data and study how 
the distribution of individual parameter influences the phase separation. 
For all simulations the desired velocity is normal distributed 
$v_i^0 \sim \mathcal{N}(\mu, \sigma^2)$ with average $\mu=1.24$ m/s and
variance $\sigma^2=0.05$. Fig. \ref{fig:simohne} and  Fig. \ref{fig:simmit}
show the simulation results for two different choices  of parameter
distributions.   

The model yields the right macroscopic relation between velocity and density
even if  $a_i, b_i, st_i $ and $\tau_i$ are the same for all pedestrians, see
Fig. \ref{fig:simohne} (left). The trajectories display that phase separation
does not appear. Even at high densities the movement is ordered and  no stops
occur. For further simulations we incorporate a certain disorder by choosing
the following individual parameter normal distributed:  $a_i \sim
\mathcal{N}(0.125, 0.1)$, $b_i \sim \mathcal{N}(0.758, 0.5)$ and  $\tau_i \sim
\mathcal{N}(1.0, 0.1)$.    

Variation of the personal parameters affects the scatter of the fundamental
diagram,  see Fig.~\ref{fig:simmit}, left. Phase separation appears in the
modeled trajectories  as in the experiment, see Fig.~\ref{fig:simmit} middle
and right. It is clearly  visible that long stop phases occur by introducing
distributed individual parameters. Then the pattern as well as the change of
the pattern from $N = 62$ to  $N = 70$ are in good agreement with the
experimental results, see Fig.~\ref{fig:traj}.  Even the phase separated
regimes match qualitatively. However, the regimes appear  more regular in the
modeled trajectories.  

\begin{figure}
\centering
\includegraphics[width=0.48\textwidth]{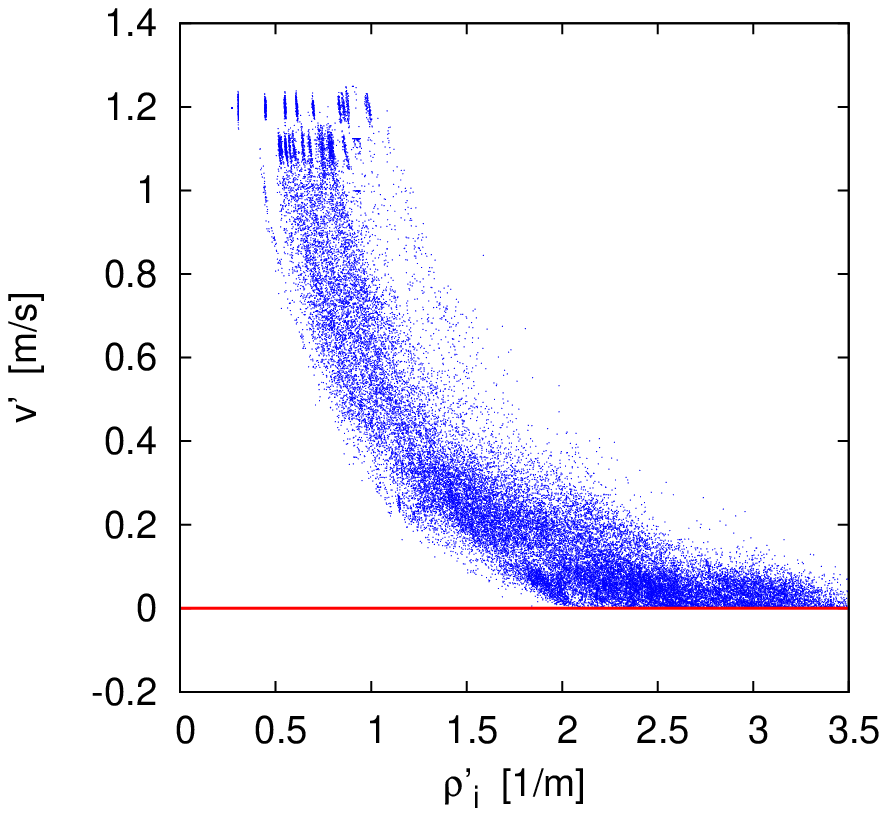}
\includegraphics[width=0.48\textwidth]{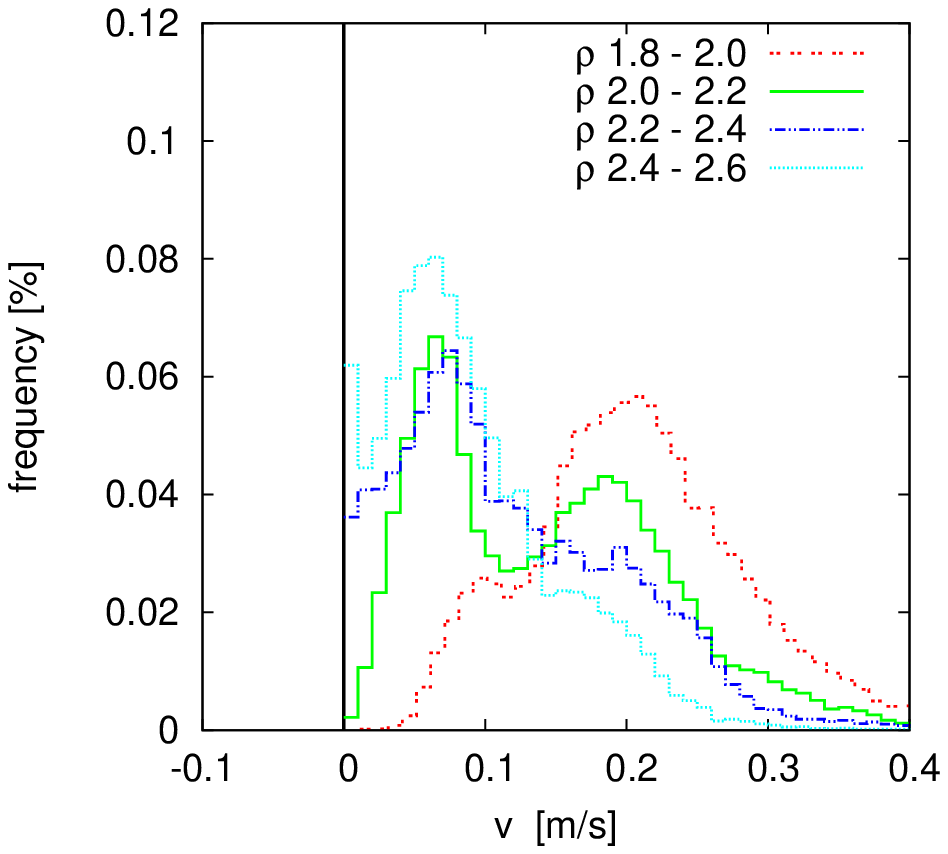}
\caption{\textbf{Left:} Microscopic fundamental diagram of the modeled data with 
normal distributed individual parameters. \textbf{Right:} Distribution of $v$ at 
fixed densities.}
\label{fig:fdmicmod}
\end{figure}

In Fig. \ref{fig:fdmicmod} microscopic measurements of fundamental diagram 
and the related velocity distributions are shown. Separation of phases is 
reproduced well. But the position of the peak attributed to the 
moving phase is not in conformance with the experimental data, compare Fig. 
\ref{fig:fdmicro} (right). The experimental data show that the peak position is 
independent from the density at $v$ around $0.15$ m/s. At the model data 
the position of the peak changes with increasing density. 
Measurements with different time steps show that the size of the time step
influences the length and shape of the stop phase at high densities. But the
density where first stops occur seems independent from the size of the time
step. Further model analysis is necessary to study the role of the reaction
time implemented by discrete time steps in this special type of
update. Furthermore we will study how the change of the peak could be
influenced by including a distribution for the step length and other
variations of the distribution for the safety distance. 

%%%%%%%%%%%%%%%%%%%%%%%%%%%%%%%%%%%%%%%%%%%%%%%%%%%%%%%%%%%%%%%%%%%%%%%%

\section{Conclusion}

We have investigated the congested regime of pedestrian traffic using
high-quality empirical data based on individual trajectories. Strong evidence
for phase separation into standing and slow moving regimes is found. The
corresponding velocity distributions show a typical two-peak structure. The
structure of the trajectories is well reproduced by an adaptive velocity model
which is a variant of force-based models in continuous space.

Future studies should clarify the origin of the differences to the phase
separated states observed in vehicular traffic. Here phase separation into a
stopping and a moving phase occurs such that the average
velocity in the moving regime is independent of the total density.  

\subsubsection*{Acknowledgments.} This work is supported by 
the Federal Ministry of Education and Research - BMBF (FKZ 13N9952 and
13N9960) and by the German Research Foundation - DFG (KL 1873/1-1 and
SE 1789/1-1).

\end{document}